\documentclass[reprint,aps,pre,superscriptaddress,nobibnotes]{revtex4-1}
\usepackage[utf8]{inputenc}
\usepackage[T1]{fontenc}

\usepackage{lineno} 

\usepackage{ microtype,enumitem,amsthm,amssymb,amsfonts,amsmath, amsbsy,amssymb,float,nicefrac}
\usepackage[stable]{footmisc}
\usepackage{listings}
\usepackage{graphicx}
\usepackage{caption}
\usepackage[]{subcaption}

\usepackage{appendix}
\usepackage[colorlinks=true]{hyperref}
\usepackage{mathtools,stackengine}

\usepackage{pgf,nicefrac}
\usepackage[textsize=scriptsize]{todonotes}

\begin{document}

\title{ANDOR and beyond: Dynamically switchable logic gates as modules for flexible information processing in biochemical regulatory networks
}

\author{Mohammadreza Bahadorian}
\affiliation{Max Planck Institut for Molecular Cell Biology and Genetics (MPI-CBG), 01307 Dresden, Germany.}
\affiliation{Center for Systems Biology Dresden (CSBD), 01307 Dresden, Germany.}
\author{Carl D. Modes}
\email{modes@mpi-cbg.de}
\affiliation{Max Planck Institut for Molecular Cell Biology and Genetics (MPI-CBG), 01307 Dresden, Germany.}
\affiliation{Center for Systems Biology Dresden (CSBD), 01307 Dresden, Germany.}
\affiliation{Cluster of Excellence, Physics of Life (POL), TU Dresden, 01307 Dresden, Germany.}

\begin{abstract}
An information processing circuit must be flexible to perform multiple tasks: biochemical information processing circuits in living systems can have multiple functions, requiring them to perform distinct computational tasks in different scenarios.  Here, we describe a tunable dynamical system that performs different logical operations (such  as AND, OR, and  XOR) depending on which basin of attraction the dynamics resides in.  We analyze the robustness of these dynamically switchable logic gates, and characterize the trade-off between reliability and efficiency in their function.
\end{abstract}

\maketitle

\thispagestyle{empty}


Biology is filled to the brim with information processing systems of all types, at all levels of complexity, and across length and time scales\cite{tkavcik2016information}. In neuronal information processing, pack animals navigate complex social decisions requiring integration of vast amounts of information about other members of the cohort\cite{black1988preflight,trillmich2004coordination}. In the realm of biochemical information processing, individual cells respond to small changes in the concentration of signaling molecules in non-linear ways via the biochemical signaling networks\cite{mora2015physical}. And the expression of genes are controlled by vast gene regulatory networks which respond to significant amounts of contextual information\cite{tkavcik2011information,tkavcik2008information}. 

The majority of efforts to understand these systems, however, conceptualize information processing as occurring over static circuits; one circuit performs one information processing task.  In order to alter this task, the circuit architecture itself or the control parameters must be changed\cite{ingram2006network}. But this is not the only possibility. In the world of silicon-based information processing, field programmable gate arrays, where the memory bits in the control layer set the connections between the logic gates in the circuit layer, are not constrained by static circuit topologies\cite{brown1992field}. Recently, neuroscience has also seen renewed attention to flexible information processing, in order to meet mounting experimental results indicating context-based task switching on timescales incompatible with plastic network adaptation\cite{vasconcelos2011cross}. 


Studying decision making in a system typically benefits from investigating the building blocks of information processing, i.e. \emph{logic gates}, in that framework. For example, discussions about quantum information processing devices have been based on their constituents: quantum logic gates \cite{barenco1995conditional}. Similarly, many studies have focused on properties of logic gates which can be implemented by biological systems \cite{matsuura2018synthetic,hasty2002engineered,katz2010enzyme}.

While traditional logic gates make a useful model of some aspects of information processing devices, they cannot render a multifunctional circuit. \emph{Dynamically switchable logic gates} are required that can switch between different functions based on demand. These circuits are particularly important in cases where the device performs multiple functions and switching occurs due to the system dynamics\cite{kirst2017shifting}. Changing the function this way enables the system to switch much faster compared to changing the structure (e.g. training)\cite{crist2001learning,scheffer1993alternative}.
Note that although some current gate designs have the capability of multifunctional operation--by increasing the output's threshold, a gate can operate as AND and OR \cite{qian2011scaling}--a systematic study of such multifunctionality is missing. Moreover, the switching here is dictated externally, directly to the output species, by changing its inhibitor concentration. This is equivalent to altering the circuit's structure, and is not a result of the underlying dynamics of the system. Therefore we here introduce some examples of \emph{dynamically} switchable logic gates, show their applicability by constructing a binary adder/subtractor, and discuss advantages and disadvantages of their multifunctionality.

That in these systems the function is affected by the dynamics could have many benefits, such as enabling switching among different tasks at a higher speed\cite{battaglia2012dynamic,kirst2016dynamic}. 
Moreover, the need for rerouting signal to a specific unit,particularly difficult in biological cases where signal is abundance of biochemical species, is circumvented.
Additionally, performing many functions with one given subsystem is crucial, especially when resources are limited. 

\begin{figure}
\centering
\includegraphics[width=.9\linewidth]{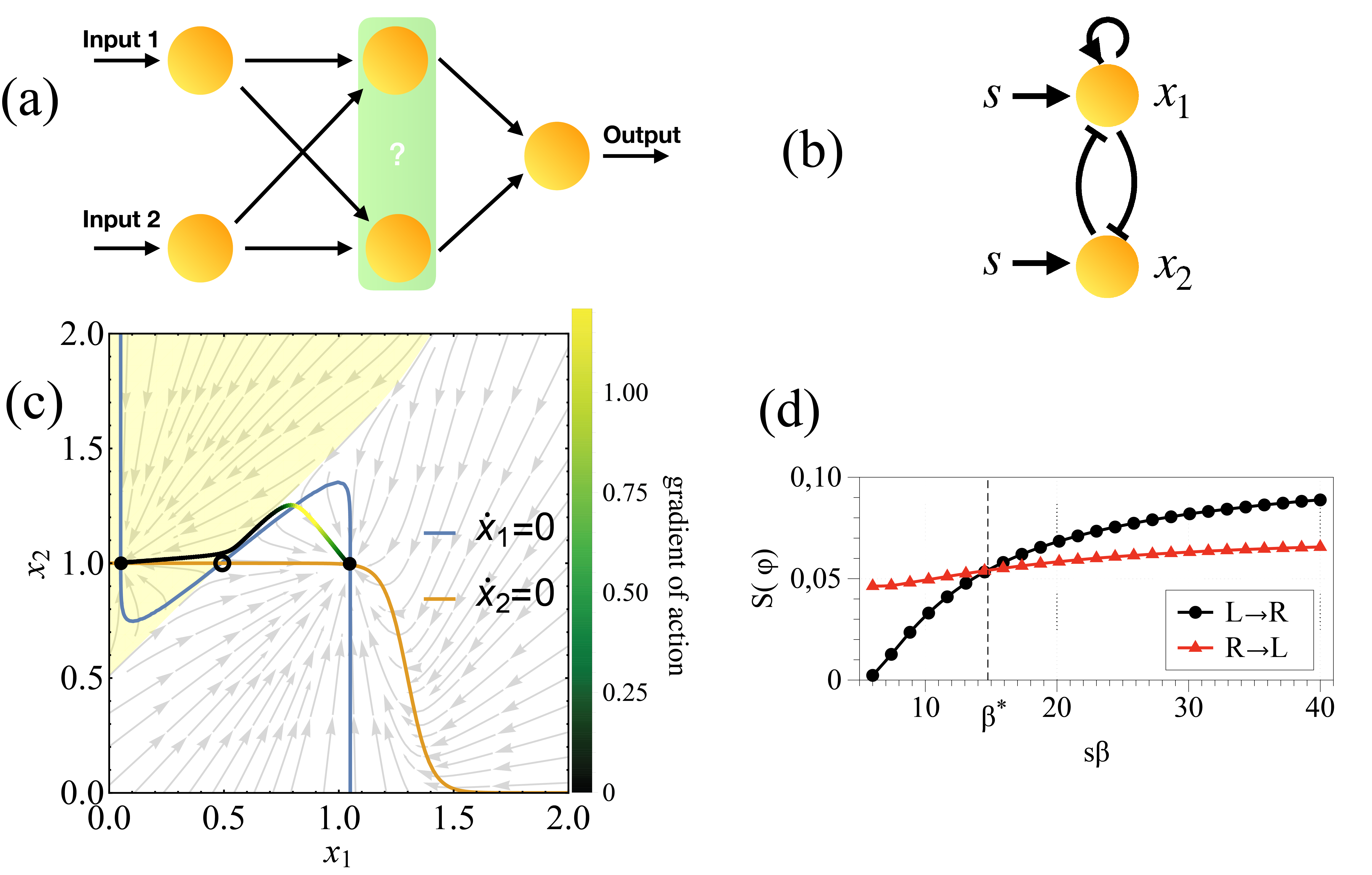}
\caption{\textbf{(a)} Generic dynamically switchable logic gate including two inputs, one output, and an intermediate layer (green box). \textbf{(b)} Topology of the motif used in the ANDOR gate.  $s\equiv \text{Input 1}+\text{Input 2}$.\textbf{(c)} Phase portrait of the bistable motif shown in (a) with intermediate signal $s=1$.  The Minimum Action Path (MAP) from middle fixed point to the left one is shown in green. The shaded area shows the basin of attraction of the left fixed point. \textbf{(d)}The minimum action for noise-induced transition vs.  the dimensionless sharpness of expression $s\beta$. \label{fig:gen_swch_gate}}
\end{figure}

We begin with a generic configuration of a dynamically switchable logic gate (Fig. \ref{fig:gen_swch_gate}a, similar to the neural networks circuit in Ref. \cite{kirst2017shifting}). Here, each element of the circuit represents a biochemical species like a gene product or other regulatory components with switch-like behavior (see App. \ref{sec:pheno_dyn}). In this configuration, an intermediate layer (green box) receives signal from two inputs (upstream genes), and sends signal to the output (downstream gene) according to its state variables. Positive signal from the two components in the intermediate layer is necessary for activating output. Throughout, links have unit weight unless stated otherwise. For simplicity, we consider the state of the left layer to be static, independent, and acquiring either zero or one corresponding to OFF and ON, respectively. We thus need only focus on the dynamics and fluctuations of the intermediate layer.

The simplest case of a dynamically switchable logic gate is one performing AND and OR functions. We call this  an ``ANDOR'' gate. By the truth tables of AND and OR, the output should be OFF regardless of the gate type when there is no input. Similarly, for both gate types, the output is ON when both inputs are ON, thus the only difference between these gates is their response to the intermediate signal level, i.e. when one of the inputs is ON and the other is OFF. In this case, the OR gate should be ON while the AND gate should be OFF. Therefore, replacing the green box in Fig. \ref{fig:gen_swch_gate}a with a sub-circuit which has bistability for intermediate input signals enables the gate to perform both OR and AND tasks (Fig. \ref{fig:gen_swch_gate}).
Here, $s$ is the sum of the two inputs to the intermediate layer which can only be zero, one or two and is assumed to be static. However, the concentration of the two genes ($x_1$, $x_2$) in the intermediate layer can have any positive real values and their dynamics are given by:
\begin{align}
\dot{x}_1&=  \frac{\lambda_0}{1+e^{-\beta\left(s-x_2+\omega\left(x_1 - \alpha_1 \right)\right)}} -\lambda_1 x_1 + \zeta \label{eq:bistbl_dyn1}  \\
\dot{x}_2&=  \frac{\lambda_0}{1+e^{-\beta\left(s-x_1 - \alpha_2 \right)}} -\lambda_2 x_2, \label{eq:bistbl_dyn2}
\end{align}
where $\omega$ controls the slope of the separatrix, and $\lambda_1$, $\lambda_2$ are the degradation rates for $x_1$, $x_2$, respectively. $\lambda_0$ is the production rate of both genes when they receive sufficiently strong input.  $\alpha_1$, $\alpha_2$ are the activation thresholds for $x_1$, $x_2$ which in our case equal to $0.5$ and $0.3$. Finally, $ \zeta \ll 1$ is needed for numerical stability. See Sec. \ref{sec:pheno_dyn} for more details.

This motif satisfies the conditions for no input and two inputs with a proper set of parameters.
Fig. \ref{fig:gen_swch_gate}c shows the phase portrait of the ANDOR gate intermediate layer along with its nullclines. These nullclines cross each other three times, producing three fixed points. Two of them are stable separated by a saddle point. The stable fixed point at the center corresponds to the situation where both genes are expressed and available for the downstream genes while the other stable fixed point corresponds to the situation in which only $x_2$ is expressed, and $x_1$ is repressed. Therefore, using this motif enables the system to perform both AND and OR functions. For only one input, if the system approaches the fixed point in the center, the output will be ON meaning that the system works as an OR gate. On the other hand, if the genes reach the left stable fixed point, the total signal for expression of the output gene is not enough and the whole circuit functions as an AND gate. 

Whether the ANDOR gate performs AND or OR functions depends on the ``context'' of the decision, and switching between these two functions is possible by transiting from one basin of attraction to the other. This transition can happen due to an additional, external signal to gene $x_2$. By such a signal, the expression of $x_2$ can be controlled and the system can be steered vertically to the desired basin of attraction. With such a strategy, cells could potentially make different decisions depending on the tissue they are part of, and the external signal they receive. 
Moreover, the decision also depends on the initial content of expressed genes. This could be employed in cell fate decision making processes where two daughter cells acquire different fates due to asymmetric distribution of cell contents during division. For example, a transcription factor of $x_2$ could be divided asymmetrically between daughter cells, leading to two distinct decisions and cell fates. 
However, if the system is obliged to start from the origin of the phase space, the final state can be controlled by adjustment of the slope of the separatrix. In this scenario, the function the system performs can be interpreted as a ``ground state'' of the system since it is the function performed naturally, without requiring injection of external energy. 

Note that the ANDOR gate only requires five components while traditional static AND and OR gates together require six. Besides, an additional controller unit incurring an even higher component cost is needed to redirect the signal to the desired gate if function switching is not possible.  Therefore, using the ANDOR gate reduces the number of required components significantly.

Although the interactions in the intermediate layer of the ANDOR gate and the resulting bistability enables the system to perform two distinct functions without requiring twice as many components, it also allows undesired noise-induced transitions (i.e. errors) that reduce the reliability of the decisions. In order to study these transitions, we minimize the Freidlin-Wentzell action to find the most probable path taken by the system for a transition, and the probability of that path up to normalization\cite{freidlin1998random}.  See Sec. \ref{sec:action_functional} for more details. 

Due to random timings of the chemical reactions in biological systems,  the concentrations of species follow stochastic dynamics. In the case of well-mixed systems, one can use the chemical Langevin equation \cite{gillespie2000chemical} to fully describe the dynamics when fluctuations are sufficiently small (i.e. the reaction volume is large). For the ANDOR gate, with dynamics described by Eq. \ref{eq:bistbl_dyn1} and \ref{eq:bistbl_dyn2}, the action over the path $\varphi$ is:
\begin{equation}
\begin{aligned}
S(\varphi) =&\frac{1}{2} \int_{T_{1}}^{T_{2}} \frac{\left(\dot{\varphi}_{1}-\left[-\lambda_1\varphi_{1}+g\left(\varphi_{1}, \varphi_{2}\right)\right]\right)^{2} }{\lambda_1\varphi_{1}+g\left(\varphi_{1}, \varphi_{2}\right)}
d t \\
&+\frac{1}{2} \int_{T_{1}}^{T_{2}} \frac{\left(\dot{\varphi}_{2}-\left[-\lambda_2 \varphi_{2}+h\left(\varphi_{1}, \varphi_{2}\right)\right]\right)^{2}}{\lambda_2 \varphi_{2}+h\left(\varphi_{1}, \varphi_{2}\right)} d t
\end{aligned}
\label{eq:acation_1}
\end{equation}
in which $\varphi_{1}$, $\varphi_{2}$ are the coordinates of the path $\varphi$ at any time $t \in \left[ T_1, T_2 \right]$.
$g\left(\varphi_{1}, \varphi_{2}\right)$ is the expression of $x_1$ and $h\left(\varphi_{1}, \varphi_{2}\right)$ is the expression of $x_2$. The probability this path is taken by the system is proportional to $ e^{-S(\varphi) / \varepsilon^2}$ where $\varepsilon$ is the noise strength, here equal to $\Omega^{-1/2}$ with $\Omega$ the reaction volume.  When $\Omega$ is large, all paths from a given point to another will have negligible probabilities compared to the one which minimizes the action in Eq. \ref{eq:acation_1}. This path, the Minimum Action Path (MAP), determines the path with the highest probability for a given transition.

The MAP which connects the central fixed point to the left one for a typical set of parameters is shown in Fig. \ref{fig:gen_swch_gate}c. The color on the path shows the gradient of the action at any given point. This gradient can be interpreted as the effective force applied by the noise causing movement along the MAP. 
The MAP goes directly towards the separatrix against the stream lines. It then follows the streamlines along the separatrix until approaching the saddle point, and finally enters the other basin of attraction, following those streamlines to the left fixed point. That the MAP for the ANDOR gate crosses the separatrix close to the saddle point is consistent with the previous findings\cite{pisarchik2014control}.

The rate of noise-induced transitions is a proxy for the reliability of the decisions. Therefore, we investigate this reliability as the parameters change (see App. \ref{app:noise_induced_transitions}).
Fig. \ref{fig:gen_swch_gate}d shows the action vs. sharpness of activation, $\beta$, for the transitions from the left fixed point to the right one and vice versa. As $\beta$ increases, the action for both transitions increase, but the L$\to$R transition increases faster resulting in a critical value at $s\beta^{*}=14.75$. Note that although higher values of $S$ mean higher reliability, any difference between the actions for two transitions results in a bias in the steady state probability distributions. The dependence of the action on the other parameters is shown in App. \ref{app:noise_induced_transitions}.
Parameters consistent with one strongly stabilized basin at the expense of the other are attainable. On the other hand these parameters could be poised at criticality in order to equalize the transition probability and minimize the bias. Rapid switching would be promoted in this scenario.

\begin{figure}
\centering
\begin{subfigure}{.25\linewidth}
\includegraphics[width=\linewidth]{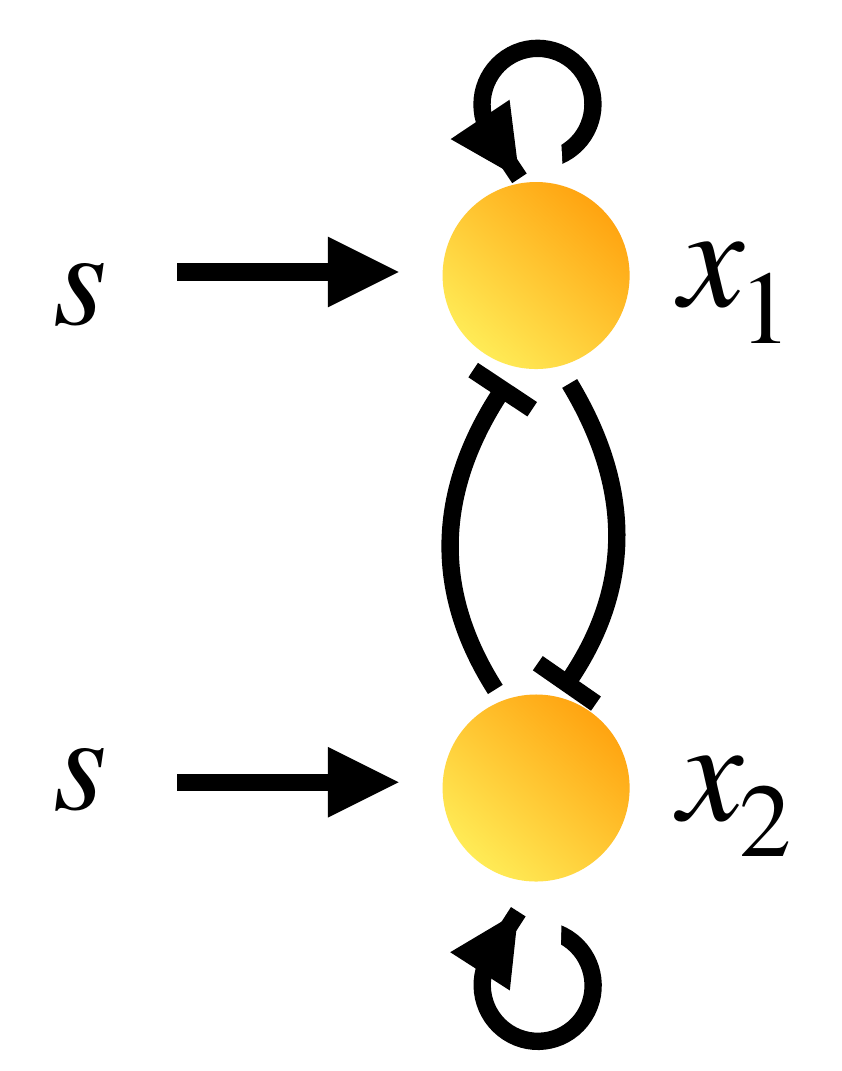}
\caption{}
 \label{fig:multistale_1}
\end{subfigure}
\hspace*{.25cm}
\begin{subfigure}{.55\linewidth}
\includegraphics[width=\linewidth]{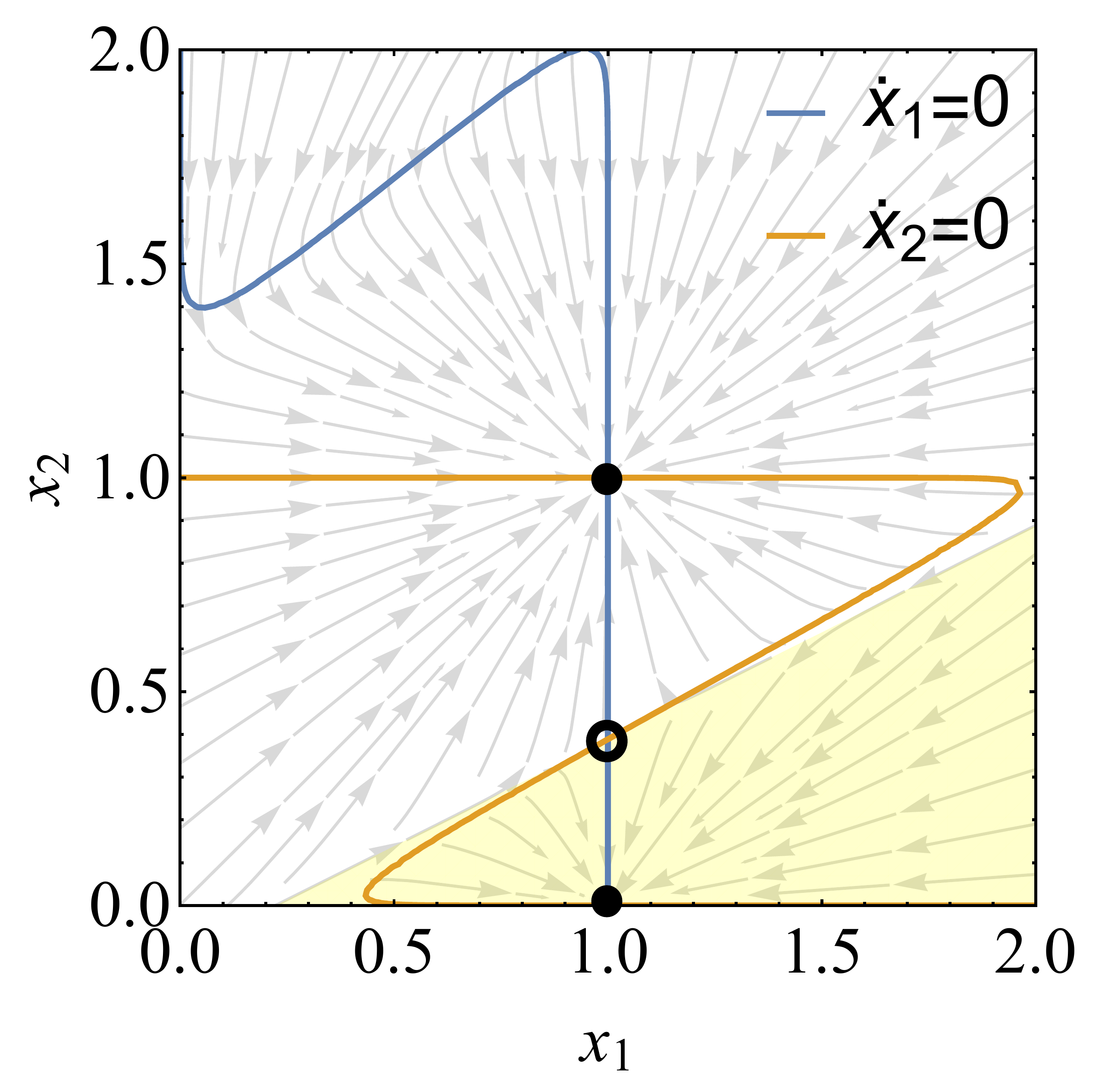}
\caption{}
\label{fig:phase_port2}
\end{subfigure}\\
\centering
\begin{subfigure}{.4\linewidth}
\includegraphics[width=\linewidth]{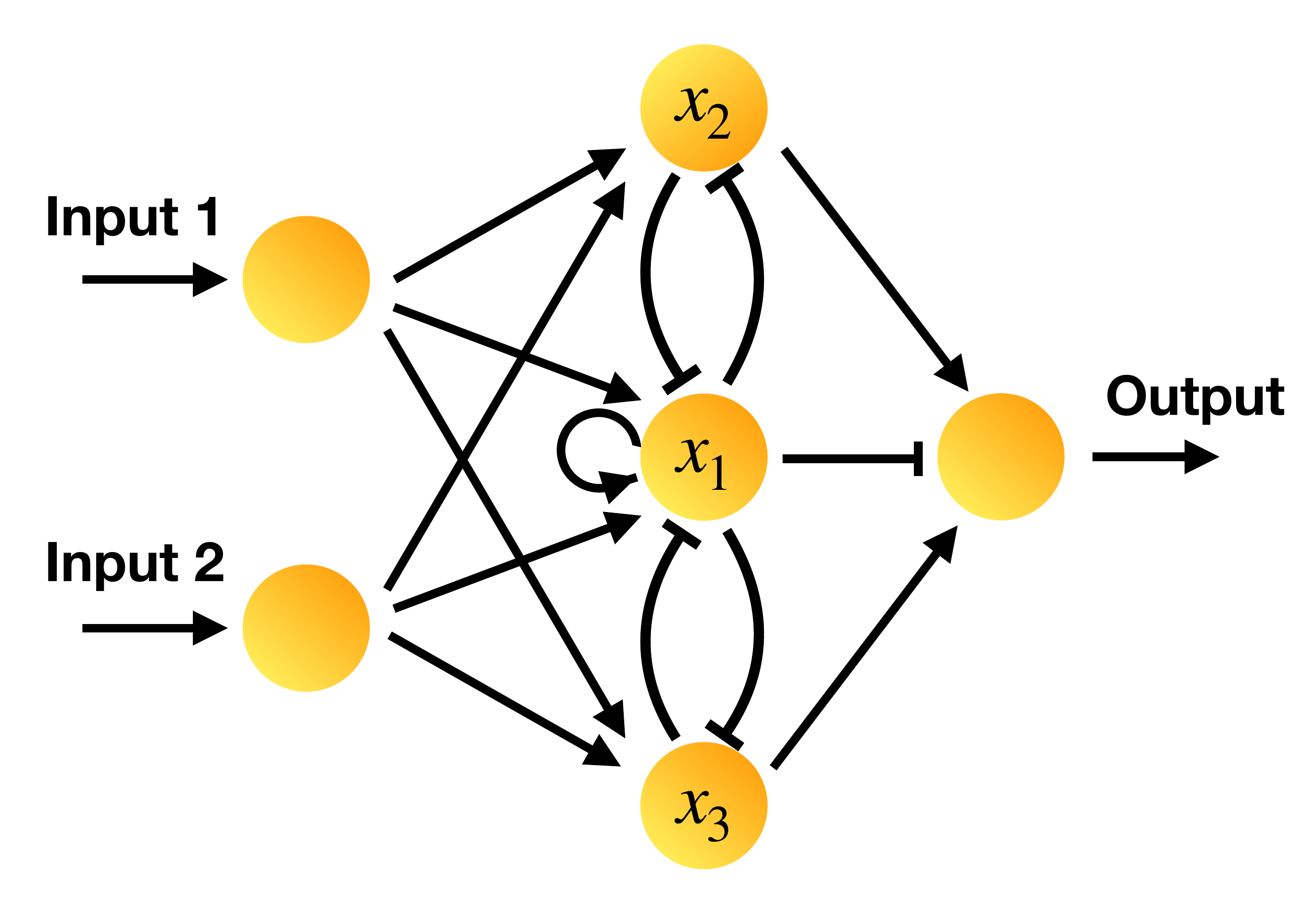}
\caption{}
\label{fig:multi_stable_circ_top}
\end{subfigure}
\hspace*{.1cm}
\begin{subfigure}{.4\linewidth}
\includegraphics[width=\linewidth]{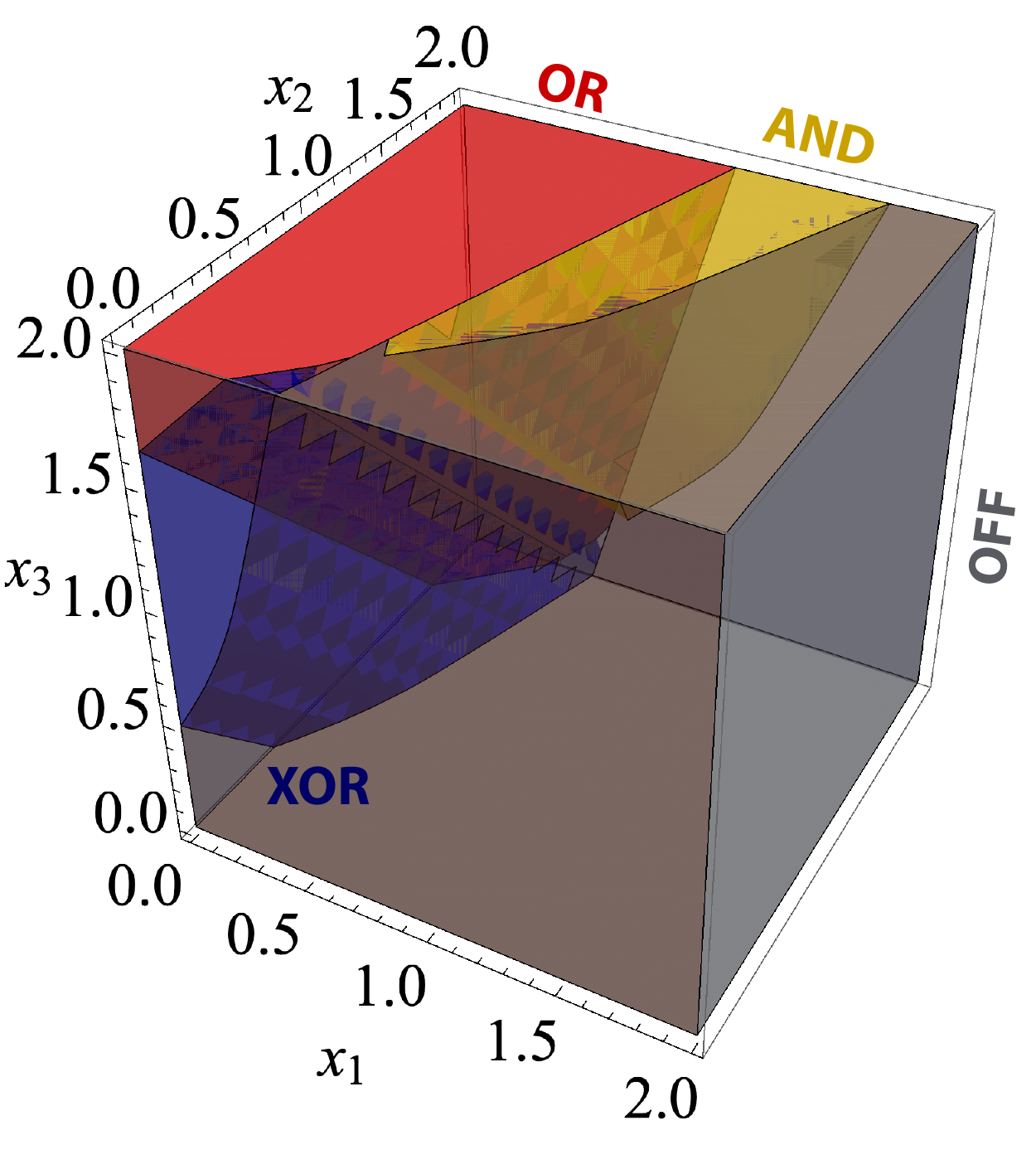}
\caption{}
\label{fig:multi_stable_circ_basin}
\end{subfigure}
\caption{\textbf{(a)} Topology of the ANDOROFF gate: a multistable motif capable of performing AND, OR and OFF.\textbf{(b)} Phase portrait of the bistable motif shown in (a) with two inputs ON ($s=2$). Shaded area shows the basin of attraction of the lower fixed point.  \textbf{(c)} The XORANDOROFF gate: a multistable circuit which can dynamically switch among XOR, AND, OR and OFF. \textbf{(d)} Regions of phase space corresponding to each function. }
\end{figure}

By increasing the complexity of the intermediate layer, one can consider circuits able to switch among a greater number of distinct functions. Consider the situation in which both genes in the toggle switch have a self-induction loop (Fig. \ref{fig:multistale_1}). For the intermediate signal level there are now three stable fixed points, two of them identical to the ones in Fig. \ref{fig:gen_swch_gate}b while the third corresponds to a state in which $x_1$ dominates and inhibits the expression of $x_2$. Moreover, when $s=2$ (i.e. both inputs are ON), one of the fixed points in which output is OFF will be preserved (as shown in Fig. \ref{fig:phase_port2}) if the self promotion loop of $x_2$ has high enough strength. The basin of attraction of this fixed point corresponds to the set of initial conditions from which the system cannot reach the central fixed point (neither with one input nor with two). Thus, the system acts as an OFF switch. Adding a new self loop, producing a new fixed point, enables the system to switch among three different functions: AND, OR and OFF. We therefore name it an ``ANDOROFF'' gate.

Although increasing the number of fixed points can enable the system to perform more functions, it decreases the size of the basin of attraction associated to each fixed point. Accordingly, the reliability of each decision against uncertainty in the initial conditions decreases. One way to overcome this limitation is by increasing the dimension of the phase-space by increasing the number of regulatory components at the intermediate layer. For example, the ``XORANDOROFF'' gate depicted in Fig. \ref{fig:multi_stable_circ_top} has three genes in this layer which, with a proper set of parameters, result in four fixed points ($(0,0,0)$, $(1,0,0)$, $(0,1,1)$ and $(1,1,1)$). However, for each input signal ($s=0,1$ or $2$), only two of them are available, and with the appropriate set of parameters, this circuit can switch between four functions: XOR, AND, OR and OFF. In Fig. \ref{fig:multi_stable_circ_basin}, consider a system initially located at the blue region. For no input the system approaches the fixed point at which no gene is expressed $(0,0,0)$ and output is OFF, and when one of the inputs is ON, the system, starting from this region, approaches the fixed point at which only $x_2$ and $x_3$ are expressed $(0,1,1)$ and therefore, the signal for output is strong enough to make it ON. However, when both inputs are ON all three genes will be expressed and expression of $x_1$ inhibits the expression of output. Therefore, if the system is initially located in the blue region it functions as an XOR gate. Similarly, locating the system in other regions enables it accordingly to perform other functions.

\begin{figure}
\centering
\begin{subfigure}{.45\linewidth}
\includegraphics[width=\linewidth]{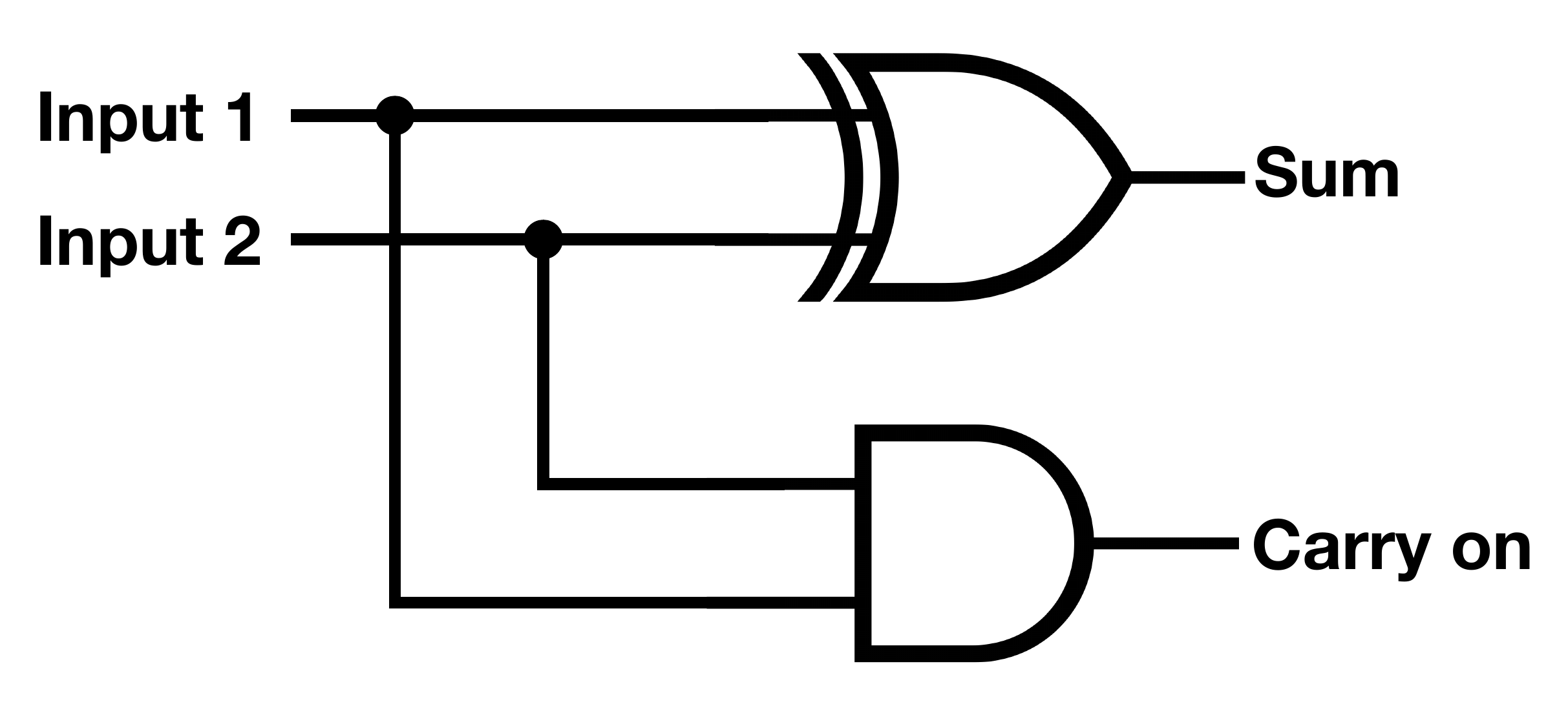}
\caption{}
\label{fig:binary_adder_circuit}
\end{subfigure}
\hspace*{.1cm}
\begin{subfigure}{.45\linewidth}
\includegraphics[width=\linewidth]{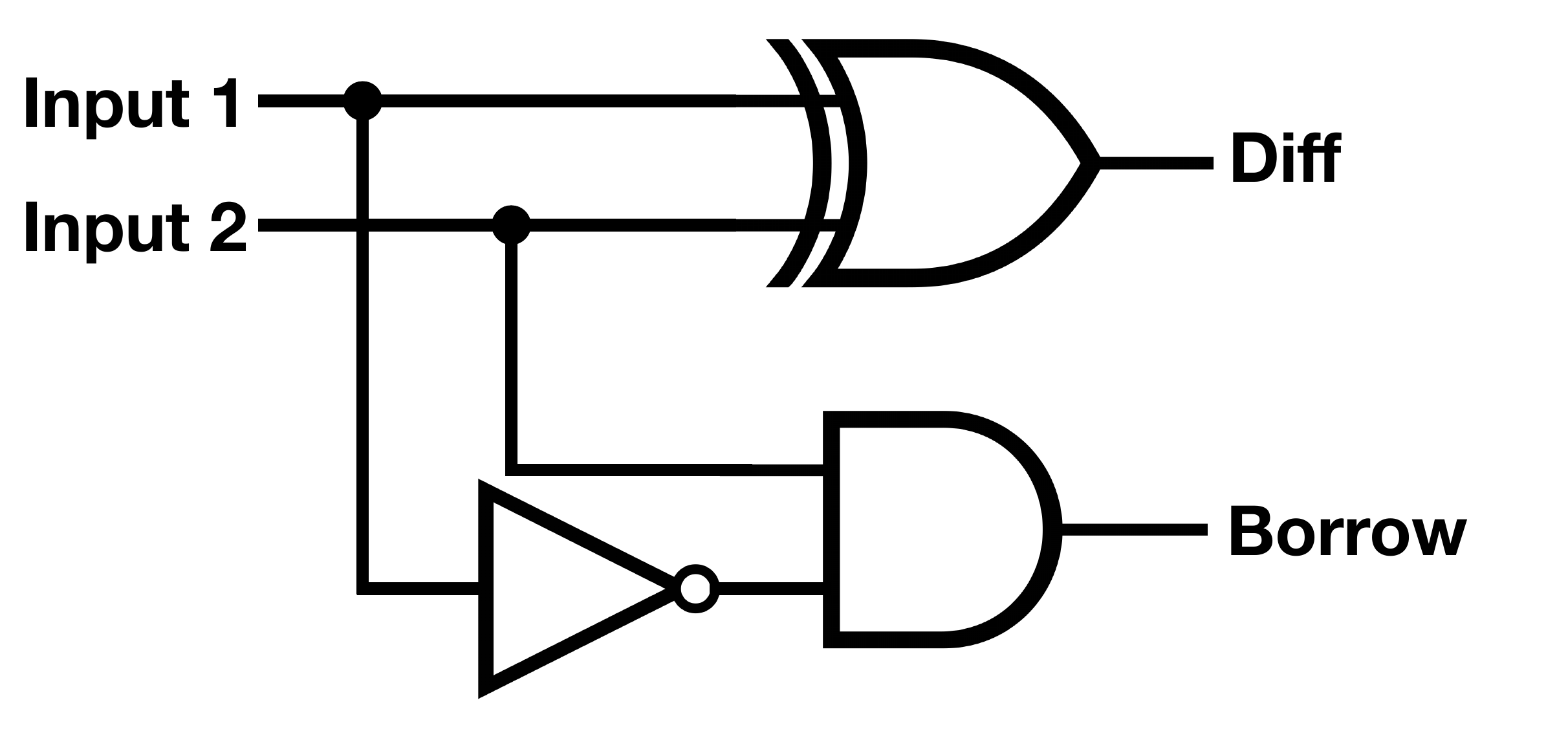}
\caption{}
\label{fig:binary_subtractor_circuit}
\end{subfigure}
\centering
\begin{subfigure}{.5\linewidth}
\includegraphics[width=\linewidth]{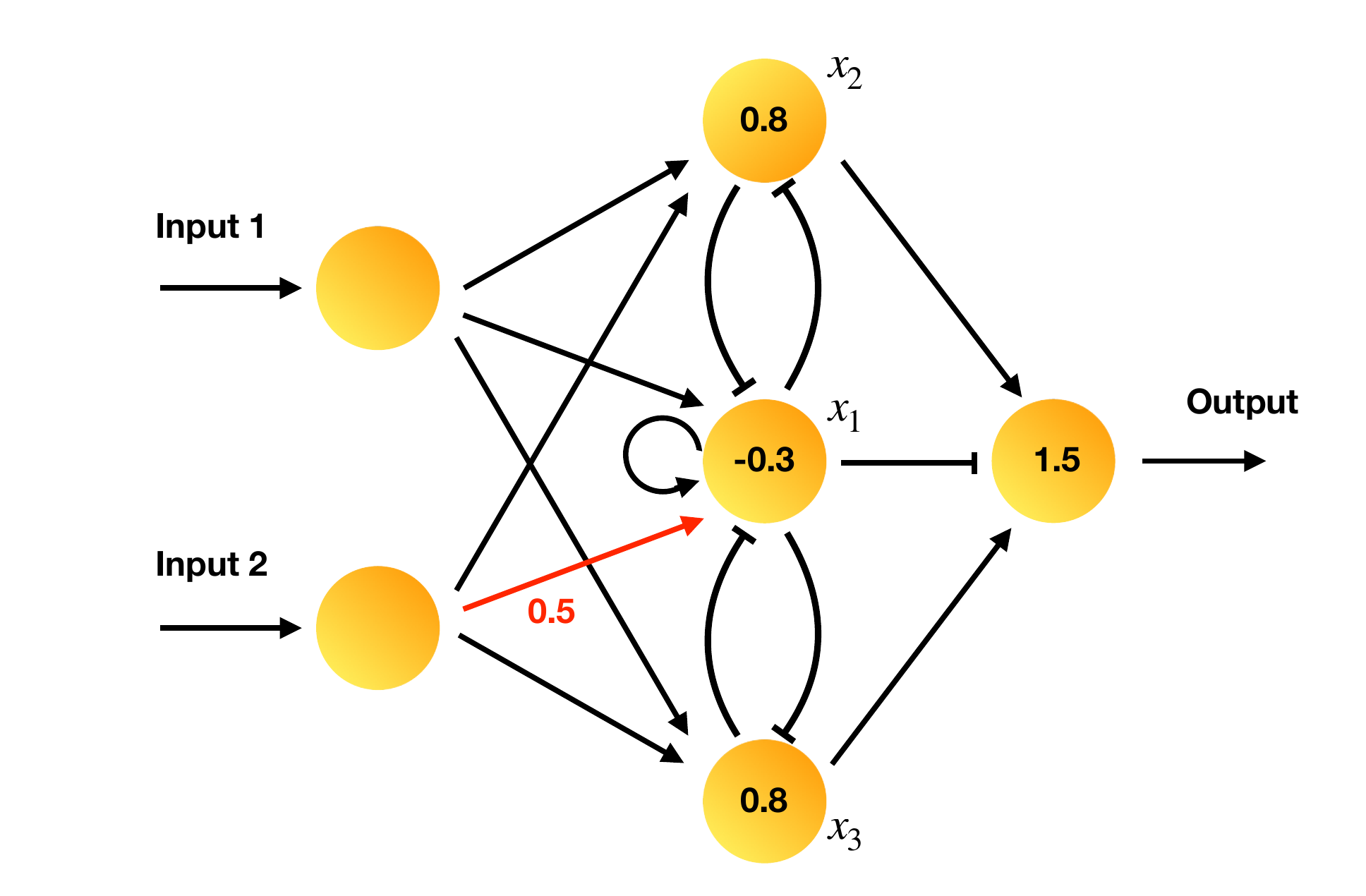}
\caption{}
\label{fig:adder_subtractor_circuit}
\end{subfigure}
\begin{subfigure}{.4\linewidth}
\includegraphics[width=\linewidth]{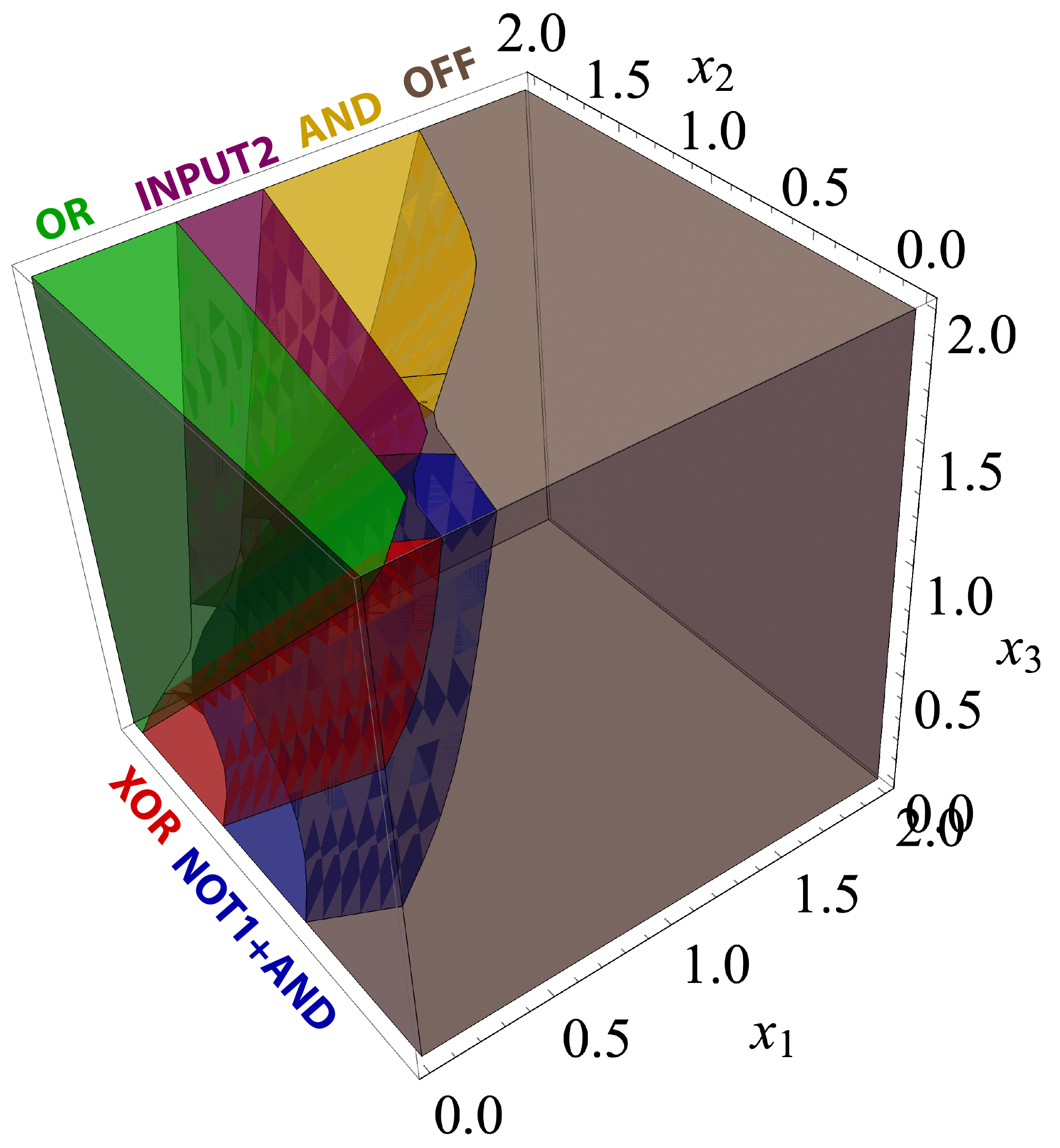}
\caption{}
\label{fig:adder_subtractor_phase}
\end{subfigure}
\caption{\textbf{Example: Adder/subtractor circuits}. Traditional binary \textbf{(a)} adder and \textbf{(b)} subtractor circuits based on static components. \textbf{(c)} Dynamically switchable binary adder/subtractor circuit.  \textbf{(d)} Initiating the system from each region shown here enables it to perform a distinct function.\label{fig:dynamical_adder_subtractor} }
\end{figure}

We demonstrate the applicability of our framework for reducing the number of required elements by designing a circuit capable of performing binary addition and binary subtraction (Fig. \ref{fig:dynamical_adder_subtractor}a,b.) \cite{sarkar2014foundation}.  

The adder circuit requires XOR and AND gates (Fig. \ref{fig:binary_adder_circuit}), while the binary subtractor needs XOR, NOT and AND(Fig. \ref{fig:binary_subtractor_circuit}). Being able to perform these two functions in a traditional static framework requires having these circuits in the system and rerouting signal (using other controller circuits) when needed. However, if we allow context-dependency, only three functions are required: AND, XOR and NOT1+AND.  We already showed having three nodes in the intermediate layer AND, XOR and more. By breaking the symmetry of the inputs to the intermediate layer and adjusting the other parameters, one may modify this circuit into the one shown in Fig. \ref{fig:adder_subtractor_circuit} capable of performing the three logic functions required for binary adder/subtractor. This circuit is also able to perform OR, OFF or reflect input 2.


We here have designed and demonstrated three examples of dynamically switchable logic gates, based on a multi-stability emerging from the interaction between components with dynamics typical to regulation of signaling biochemical species. These gates can switch between two, three and four distinct functions. We investigated the parameter ranges in which the conditions for this multi-functionality are met. Using the theory of large deviations, we characterized noise induced transitions between the possible logical functions and determined the reliability of the decisions. We therefore demonstrated that the proposed dynamical logic gates are resilient against three types of uncertainty: uncertainty in initial conditions, in control parameters, and in dynamics. 

Looking forward, one observes that existence of memory is a necessary feature for constructing high-order information processing circuits. From an engineering perspective, sequential logic circuits are employed in which the output depends not only on the current input but also on the history of inputs. In the context of biological information processing,this effect could be achieved, by addition of a toggle switch to an existing combinational (i.e. memoryless) logic gate \cite{lou2010synthesizing}. In our framework, however, the bistability that results in switchability may also play a dual role, providing a memory of the last action without requiring an extra toggle.

We believe that these are only the first steps towards a more complete and faithful understanding of how network topology, dynamical systems, and information processing can combine in powerful, flexible, and non-trivial ways in the context of gene regulatory networks, signaling networks, and chemical computations. Many questions and implications remain to be explored. Thus far we have considered these biochemical dynamics playing out under the assumption of a well-mixed reaction volume, but spatial localisation and compartmentalisation are more and more appreciated as being important players in cell biology\cite{banani2017biomolecular,gonzales2020building} and could have significant effects on the operation of these switchable information processing elements. Another important direction will be to consider how families of more complex calculations built from the ANDOR gate and its cousins can be coupled to adaptive pressures and response on evolutionary timescales. It seems plausible that restrictions on total chemical signaling species number could make building different computations out of the same elements an attractive possibility. Finally, of course, the search is on for in vivo examples of ANDOR gates. 

Beyond the direct follow up questions, though, deeper implications also present themselves. Could one take advantage of these switchable gates to build an analog, chemical computer version of a deep learning network? Might biology have already done something similar? One might even imagine that chemical computations made flexibly switchable by simple but non-trivial combinations of network topology and reaction dynamics could have played a key role in the initiation of adaptation and natural selection at the origin of life.





\section*{Author contributions statement}

This study is designed and conducted by M.B. and C.M.. The numerical work is done by M.B. and the manuscript is written and revised by M.B and C.M.

\onecolumngrid
\newpage
\appendix

\section{ regulatory dynamics \label{sec:pheno_dyn}}
We here review general regulatory dynamics in biological systems. The time evolution of the concentration of the expressed gene $x$ is governed by a simple, nonlinear dynamical equation composed of two terms corresponding to the production $F\left(s\right)$ and degradation $\lambda x$, i.e.
\begin{align}
\dot{x}(t)=F\left(s(t)\right)-\lambda x(t),
\label{eq:dyn_gen}
\end{align}
when the transcription process is faster than translation \cite{panovska2013gene}. Here, $\lambda$ is the degradation rate, $s$ is the sum of all incoming regulatory signals (inductive, inhibitory, and self-regulation loops) to this gene, and $F\left(s\right)$ is the regulatory function that describes the response of the expression rate to the input regulatory signal $s$. Throughout this study, we use a phenomenological and commonly used sigmoidal function
\begin{align}
F\left(x,s\right)=\frac{1}{1+e^{-\beta\left(s-\alpha\right)}}~~,
\label{eq:expression}
\end{align}
where $\beta$ controls the sharpness (i.e. inverse of the fuzziness), and $\alpha$ controls the location of sigmoid function or in other words, the threshold in the signal after which the gene will be expressed \cite{jimenez2017spectrum}. It should be noted that the threshold $\alpha$ can also acquire negative values which means that the gene will be expressed even if it is inhibited with a strength smaller than this negative value. 

\begin{figure}
\centering
\begin{subfigure}{.3\linewidth}
\includegraphics[width=\linewidth]{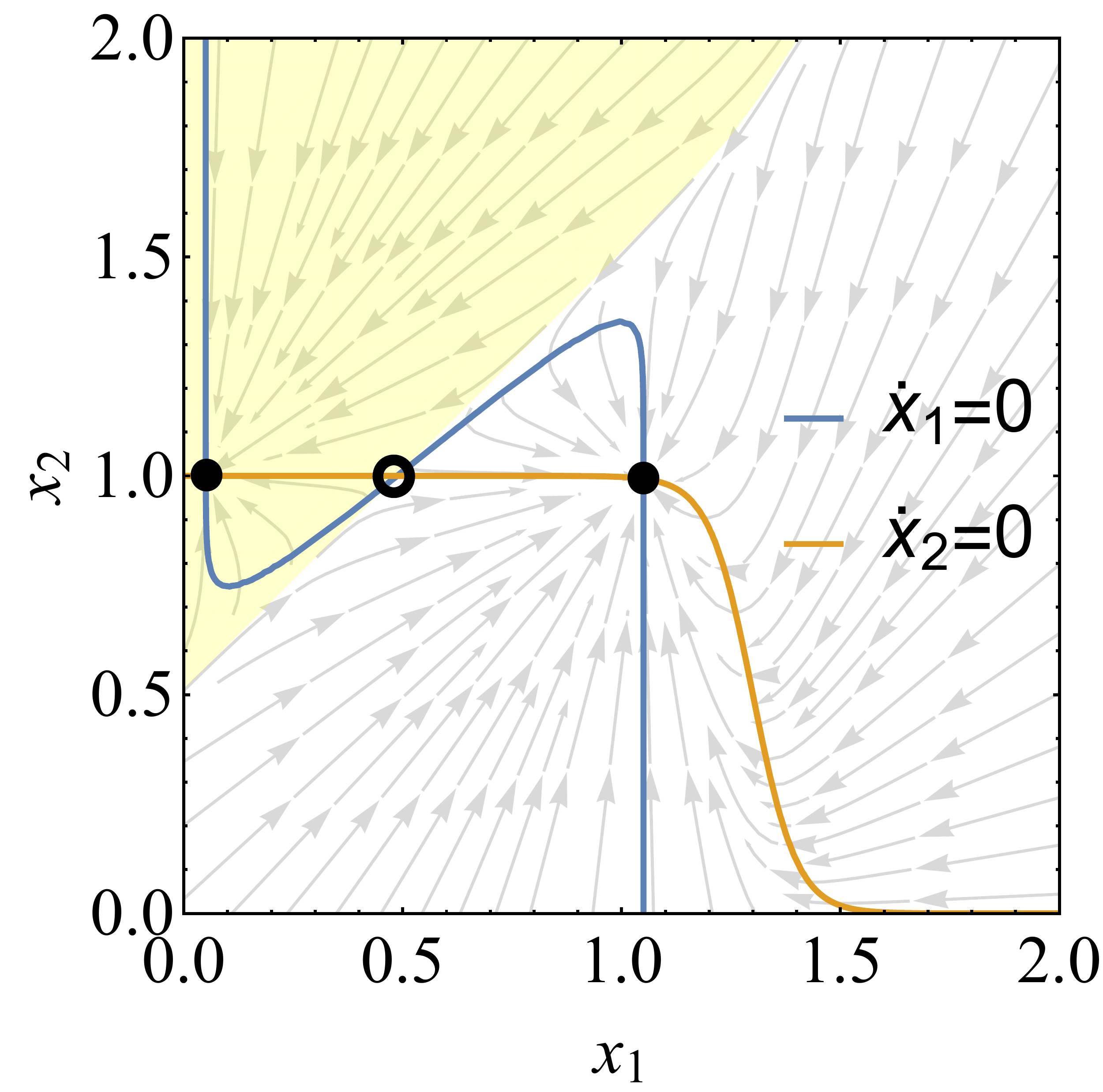}
\caption{}
\label{fig:phs_prt_phen}
\end{subfigure}
\begin{subfigure}{.3\linewidth}
\includegraphics[width=\linewidth]{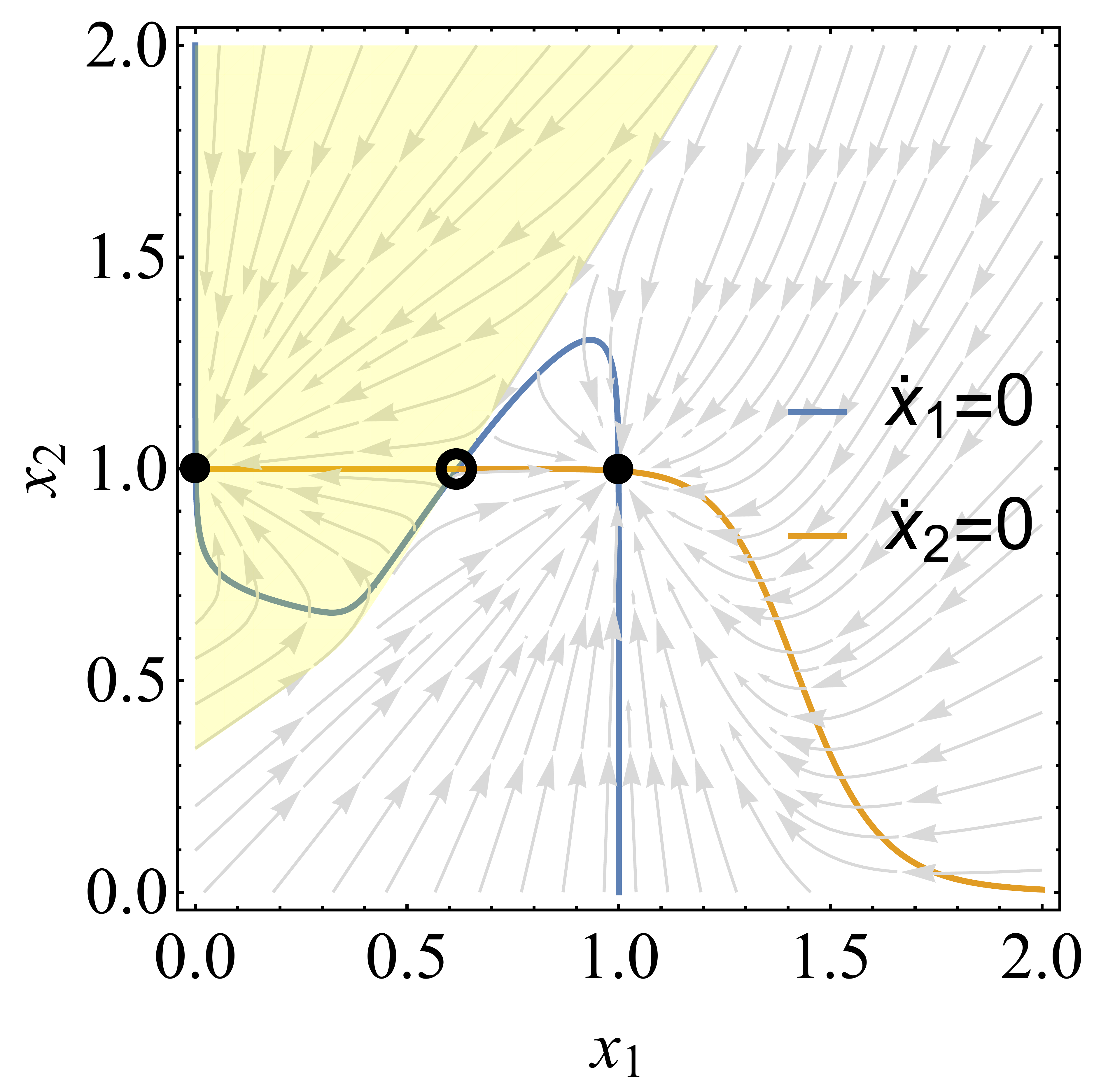}
\caption{}
\label{fig:phs_prt_hill}
\end{subfigure}
\caption{\textbf{Different underlying regulatory dynamics display the same bistability.} The phase portraits of the bistable motif in Fig. \ref{fig:gen_swch_gate}c \textbf{(a)} with the phenomenological dynamics and in \textbf{(b)} with Hill function dynamics. In both of these plots, the solid black circles show the fixed points, the hollow ones show the saddle points and the shaded area show the basins of attraction of the left fixed points.\label{fig:hill_vs_phen}}
\end{figure}

Although we are using a specific type of dynamics for our examples which is shown in Eq. \ref{eq:expression}, one can in principle use any other type of regulatory function (e.g. the Hill function) as long as it features a switch-like sigmodal behavior and dynamically switching should still be achievable. In order to demonstrate this, we also constructed another version of the ANDOR gate based on the Hill function regulatory dynamics instead of the one shown in Eqs. \ref{eq:bistbl_dyn1} and \ref{eq:bistbl_dyn2}: 
\begin{align}
\dot{x}_1&=   \frac{\left(\nicefrac{x_1}{0.3}\right)^n+\left(\nicefrac{s}{0.8}\right)^n}{1+\left(\nicefrac{x_1}{0.3}\right)^n+\left(\nicefrac{s}{0.8}\right)^n+\left(\nicefrac{x_2}{0.5}\right)^n} - x_1, \label{eq:hill_dyn1}  \\
\dot{x}_2&=  \frac{\left(\nicefrac{s}{0.35}\right)^n}{1+\left(\nicefrac{s}{0.35}\right)^n+\left(\nicefrac{x_1}{0.5}\right)^n} -x_2. \label{eq:hill_dyn2}
\end{align}
Assuming this dynamics for the intermediate layer of the ANDOR gate with $n=15$ results in a phase portrait that is qualitativey similar to that of Eqs. \ref{eq:bistbl_dyn1} and \ref{eq:bistbl_dyn2}. In Fig. \ref{fig:hill_vs_phen}, one can see these two plots side by side for the intermediate input signal level $s=1$.

\section{Characterization of the ANDOR gate \label{sec:andor_characterization}}
\begin{figure}
\centering
\begin{subfigure}{.2\linewidth}
\centering
\includegraphics[width=\linewidth]{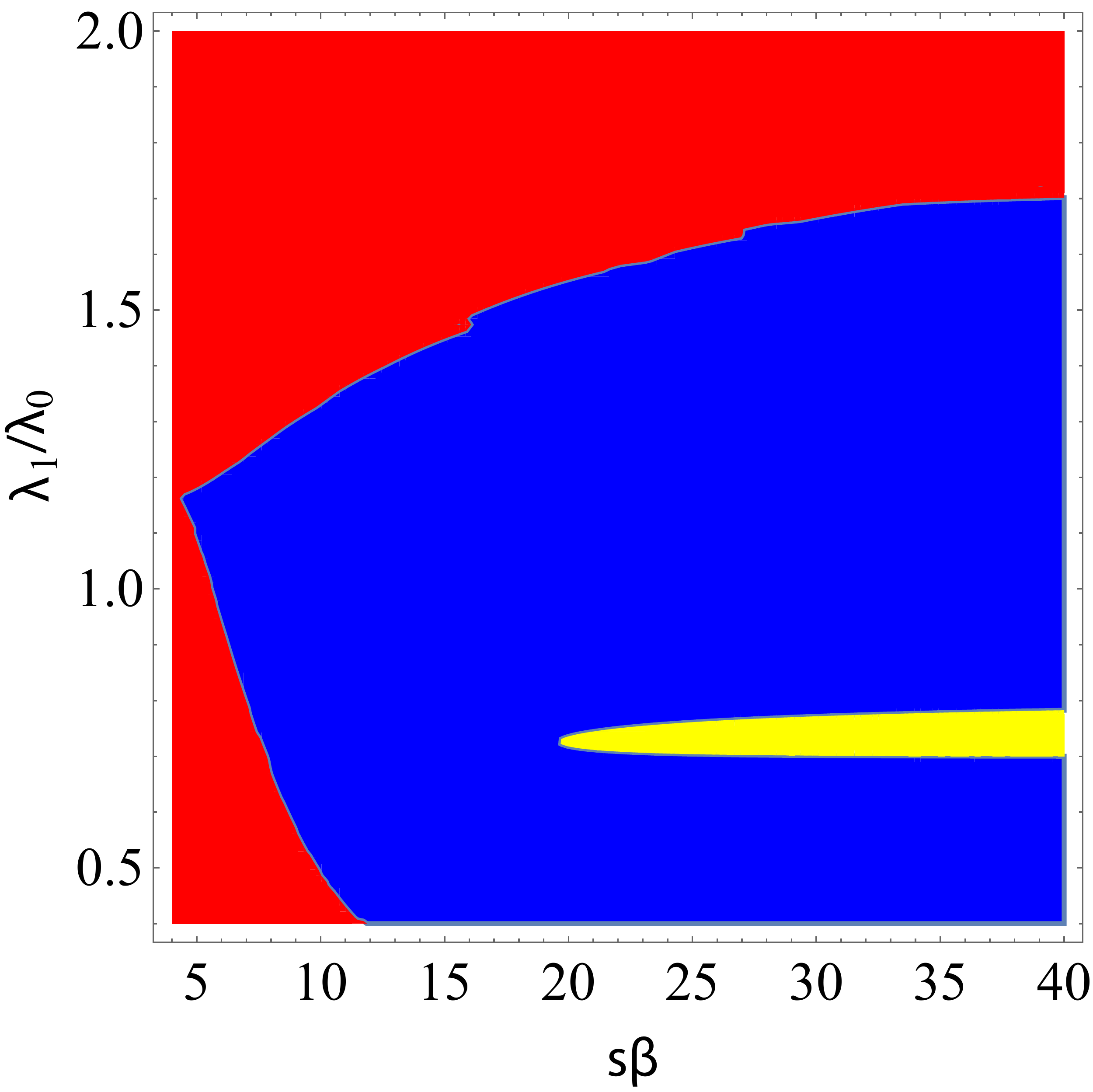}
\caption{\label{fig:phaseplot_betalambda1} }
\end{subfigure}
\begin{subfigure}{.2\linewidth}
\centering
\includegraphics[width=\linewidth]{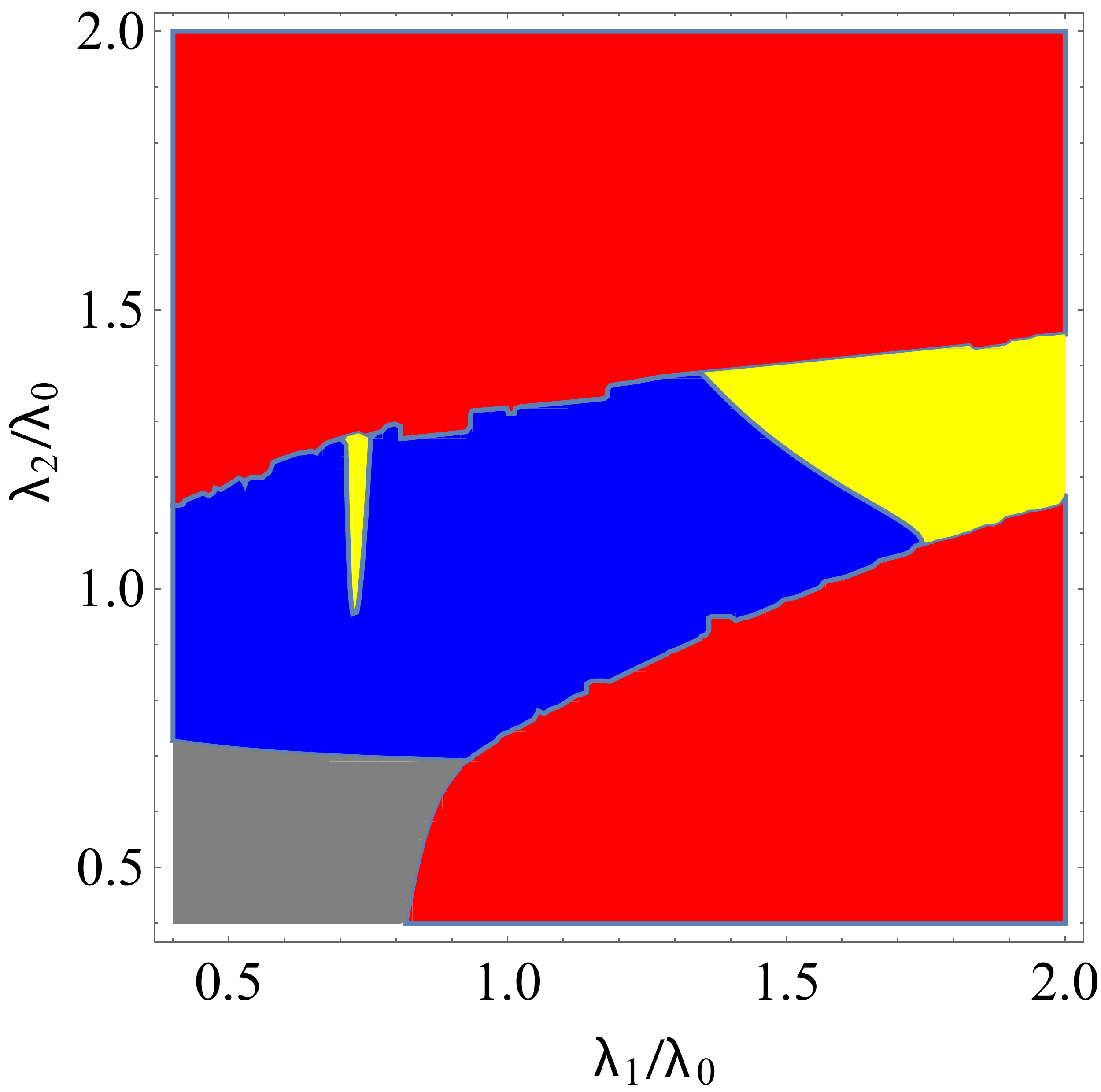}
\caption{\label{fig:phaseplot_lambda12} }
\end{subfigure}
\caption{\textbf{Phase diagram of the ANDOR gate} described by Eqs. \ref{eq:bistbl_dyn1} and \ref{eq:bistbl_dyn2} in \textbf{(a)} $s\beta$--$~\nicefrac{\lambda_1}{\lambda_0}$ and \textbf{(b)}$\nicefrac{\lambda_1}{\lambda_0}$--~$\nicefrac{\lambda_2}{\lambda_0}$ planes. Blue color shows the region in which system meets all requirements while red area shows where it does not have bistability. When the bistability exists, yellow shows where the sum of signals to the output does not meet its threshold, and gray shows where the output turns ON without any input (i.e. $S=0$). Note that in both phase plots the parameter range for which the network operates as an ANDOR gate is significant.}
\end{figure}

In order to understand the potential application of this multi-functionality, one needs to explore the phase diagrams of the system. Fig. \ref{fig:phaseplot_betalambda1} shows the different regions in the dimensionless parameter space of the $s\beta$ and $\frac{\lambda_1}{\lambda_0}$ plane when $\frac{\lambda_2}{\lambda_0}$ is set to 1. The blue area shows the parameter combinations which satisfy all conditions required for performing AND and OR functions. The yellow part shows the region in which the system at the central fixed point does not send enough signal to the output (i.e. $x_1+x_2 < 1.5$) which means that the system can not perform the AND function. Finally, the red region is where the bistability does not exist. Similarly, we also determine the right combination of $\frac{\lambda_1}{\lambda_0}$ and $\frac{\lambda_2}{\lambda_0}$ , as shown in Fig. \ref{fig:phaseplot_lambda12} for $s\beta=20$. Here, each color has the same meaning as as in Fig. \ref{fig:phaseplot_betalambda1},and the gray color shows the region where degradation for both $x_1$ and $x_2$ is so low that even without input, their steady-state concentrations meet the output threshold and it turns ON. As one can see in these figures, there is a robust range of parameters for which this circuit behaves as a context-dependent logic gate switching between AND and OR.

\section{Noise induced transitions in dynamical systems \label{sec:action_functional}}
One can define the dynamics of a stochastic system by a Langevin equation, i.e.
\begin{equation}
\dot{X}_{t}=b\left(X_{t}\right)+\varepsilon \sigma\left(X_{t}\right) \dot{W}_{t}
\label{eq:gen_stoch_dyn}
\end{equation}
where the $n$-dimensional vector $X_t$ is the state variable at time $t$, $b\left(X_{t}\right)$ is the deterministic part of the dynamics (i.e. drift vector), and the vector $W_{t}$ is a $m$-dimensional Wiener process. Moreover, $\varepsilon$ is a small parameter determining the noise strength, and $\sigma\left(X_{t}\right)$ is an $n\times m$ matrix, known as the diffusion matrix which determines the standard deviation of the noises and their contributions to each component of the dynamics. For such systems, the Freidlin-Wentzell action can be written as
\begin{equation}
S(\varphi)=\frac{1}{2} \int \sum_{i, j} a_{i j}\left(\varphi_{t}\right)\left(\dot{\varphi}_{t}^{i}-b^{i}\left(\varphi_{t}\right)\right)\left(\dot{\varphi}_{t}^{j}-b^{j}\left(\varphi_{t}\right)\right) d t
\label{eq:action_functional}
\end{equation}
where $\left(a_{i j}(x)\right)=\left(\sigma(x) \sigma^{*}(x)\right)^{-1}$. This action determines the difficulty of taking the path $\varphi_{t}$ by the system. The probability for this path to be taken at a given noise strength $\varepsilon$ is proportional to $ e^{-S(\varphi) / \varepsilon^2}$. Obviously, for small noise strengths $\varepsilon$, all paths have negligible probability compared to the one which minimizes action $S(\varphi)$.

Minimizing the action in Eq. \ref{eq:action_functional} over the function space containing all paths which connect the point $X_0$ at time $t=0$ to $X_T$ at time $t=T$ provides the Minimum Action Path (MAP), and then, the minimum action value can be used for calculation of the rate of that transition. In order to find the transition probability from one state to another, one may intuitively expect to do this procedure for the transitions from every point in one basin of attraction to the fixed point of the other basin. However, it has been shown that all trajectories which leave a basin of attraction due to the fluctuations will visit a small neighborhood around the fixed point before leaving the basin \cite{freidlin1998random}.
Therefore,  in the case of small fluctuations, the transition from one fixed point to the other represents the dominant transition and suffices for determining the resilience (against noise) of approaching the desired fixed point when starting from any point in its basin of attraction.
We thus use this measure for studying the reliability of the decisions of our gate.  It should be noted that when the system undergoes a transition from one fixed point to another, it spends most of the time at the fixed points and a small fraction of total time will be spent on the actual transition. Therefore, in order to get an acceptable accuracy, one needs to use an adaptive minimum action path method in which the distance (i.e. meshing) of time points is adaptively adjusted based on the speed \cite{zhou2008adaptive}.

\subsection{Noise induced transitions in the ANDOR gate \label{app:noise_induced_transitions}}
\begin{figure}
\centering
\begin{subfigure}{.4\linewidth}
\includegraphics[width=\linewidth]{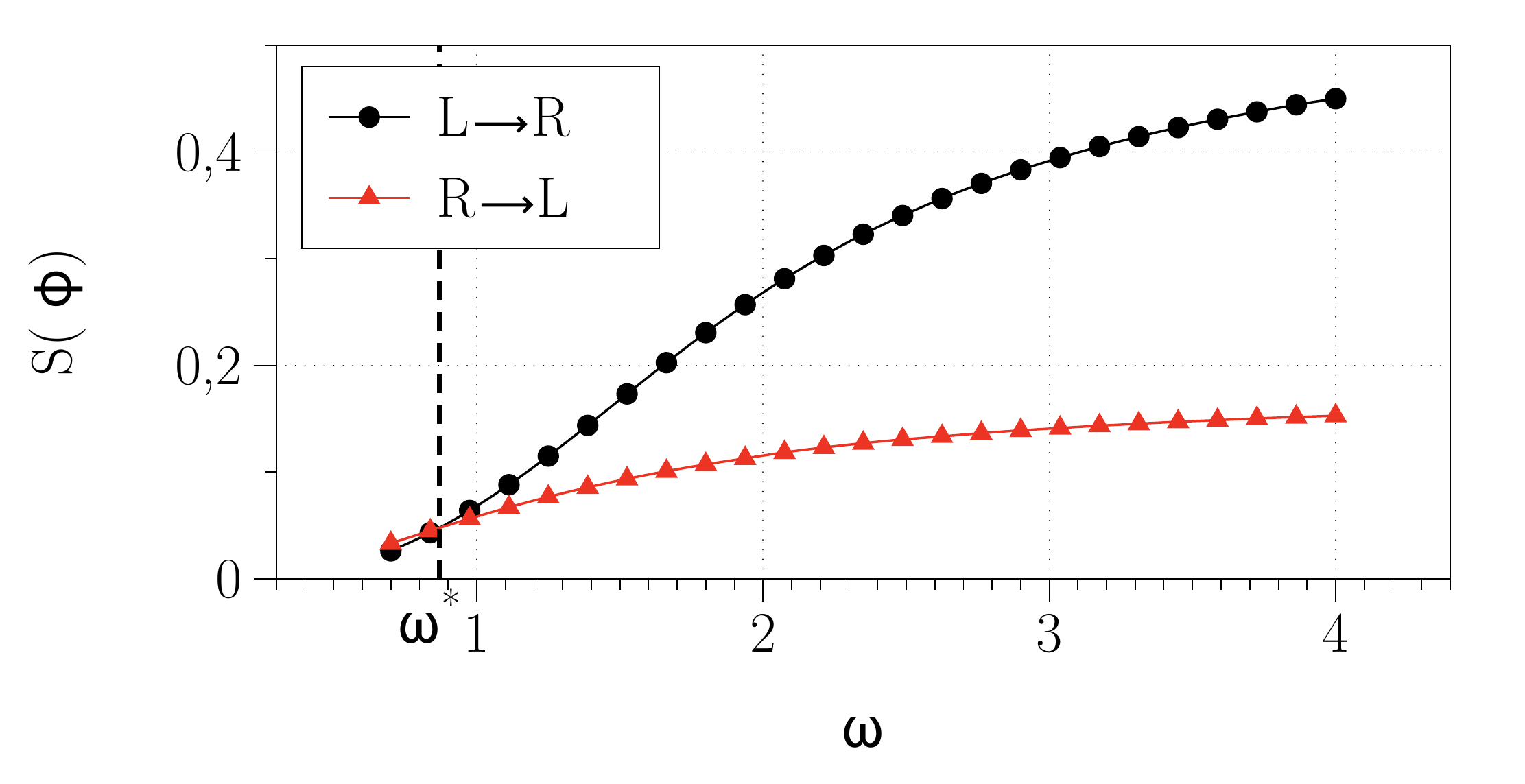}
\caption{}
\label{fig:action_vs_omega}
\end{subfigure}
\begin{subfigure}{.4\linewidth}
\includegraphics[width=\linewidth]{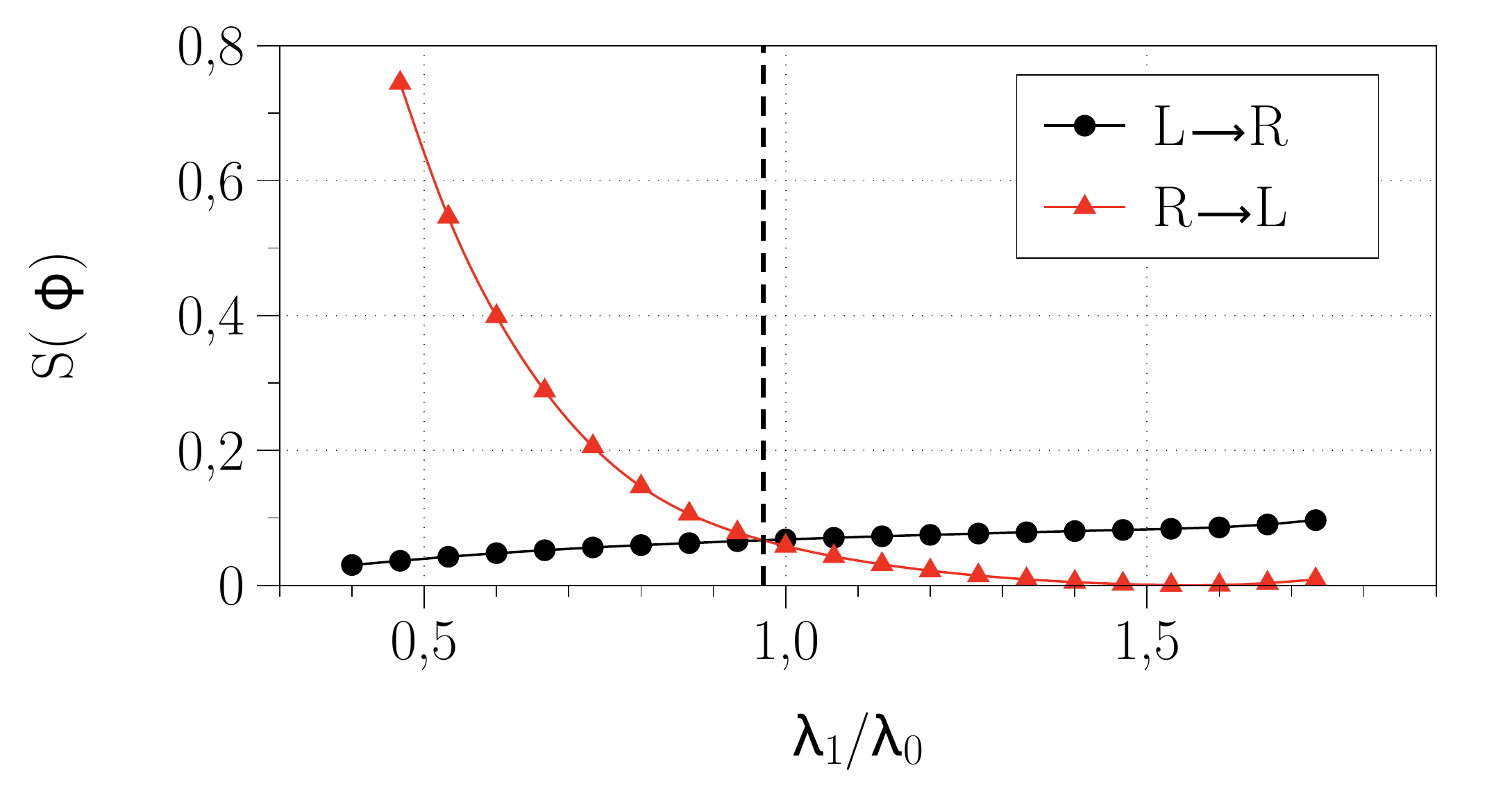}
\caption{}
\label{fig:action_vs_lambda}
\end{subfigure}
\begin{subfigure}{.4\linewidth}
\includegraphics[width=\linewidth]{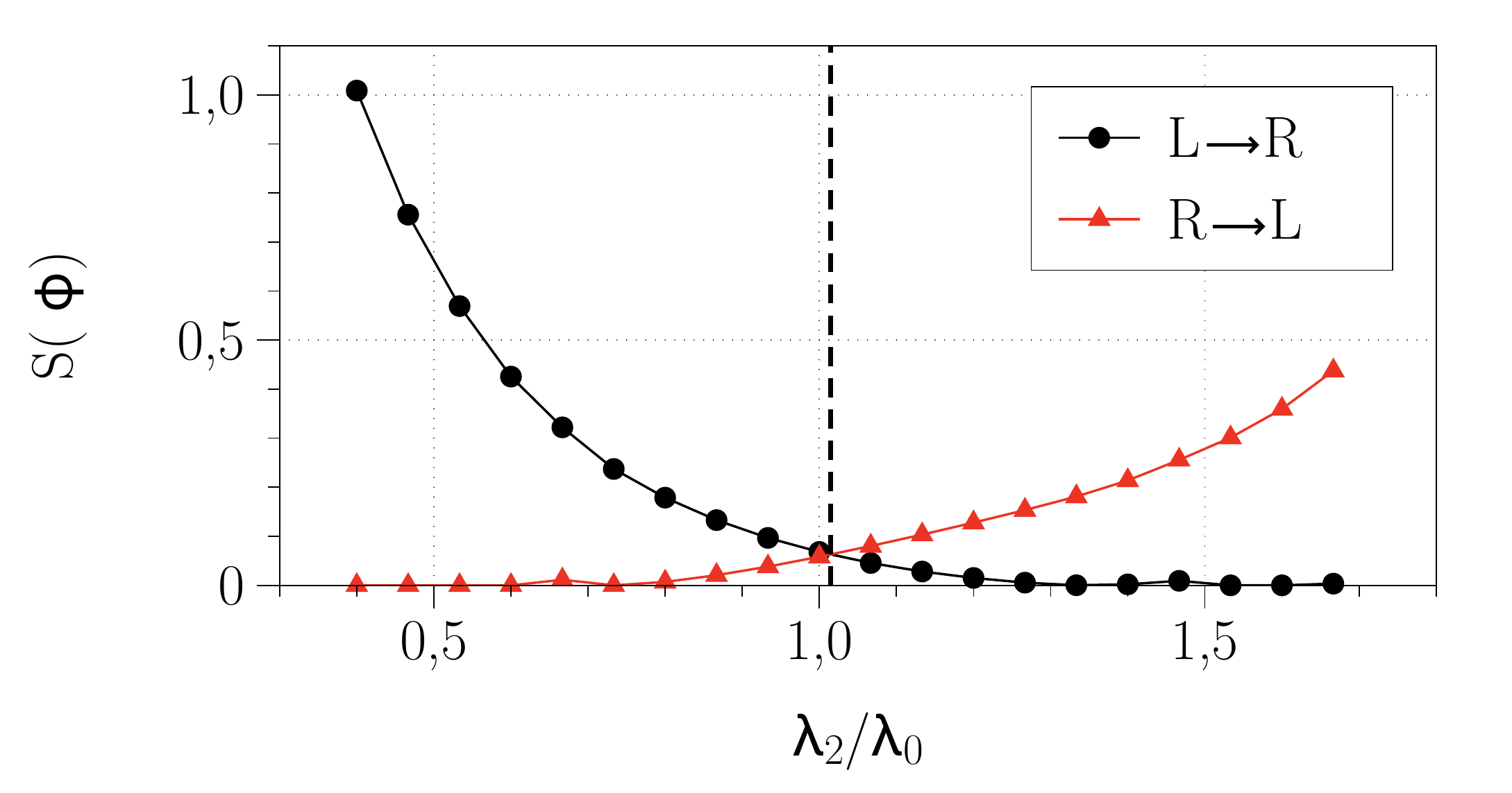}
\caption{}
\label{fig:action_vs_lambda2}
\end{subfigure}
\caption{\textbf{The minimum action for noise-induced transition} vs. \textbf{(a)} the slope of separatrix $\omega$, \textbf{(b)} the degradation rate $\lambda_1/\lambda_0$, and \textbf{(c)} degradation rate $\lambda_2/\lambda_0$.  Black circles show the action for the transition from the right fixed point the left one and the red triangles show the action for the opposite transitions.  For each dependency a critical value at which the cost of transiting L$\to$R and R$\to$L is balanced can be attained and is indicated by dashed line and have the following values: $s\beta^*=14.75$, $\omega^*=0.87$, $\lambda_1^*/\lambda_0=0.97$ and $\lambda_2^*/\lambda_0=1.01$. }
\end{figure}

Here, we present the result of studying the relation between noise induced transitions and parameters of the ANDOR gate. As mentioned in the main text, the action $S\left(\varphi\right)$ for a path connecting to fixed point can be used as a measure for the reliability of the logical functions in our system, since the rate of undesired transitions is proportional to $e^{-S\left(\varphi\right)/\varepsilon^2}$. The action vs. the slope of the sepratrix $\omega$, degradation rate $\lambda_1$ of $x_1$, and that of $x_2$ are shown in \ref{fig:action_vs_omega}, \ref{fig:action_vs_lambda},  and \ref{fig:action_vs_lambda2}, respectively.  The critical values at which the action for L$\to$R transition equals R$\to$L are $\omega^*=0.87$, $\lambda_1^*/\lambda_0=0.97$ and $\lambda_2^*/\lambda_0=1.01$. Note that in all these figures, unless a parameter is the variable of the plot we set them as following: $s\beta=20$ and $\lambda_1/\lambda_0=\lambda_2/\lambda_0=\omega=1$. 

\newpage


\end{document}